\documentclass[pdflatex,sn-mathphys-num]{sn-jnl}

\usepackage{graphicx}%
\usepackage{multirow}%
\usepackage{amsmath,amssymb,amsfonts}%
\usepackage{amsthm}%
\usepackage{mathrsfs}%
\usepackage[title]{appendix}%
\usepackage{xcolor}%
\usepackage{textcomp}%
\usepackage{manyfoot}%
\usepackage{booktabs}%
\usepackage{algorithm}%
\usepackage{algorithmicx}%
\usepackage{algpseudocode}%
\usepackage{listings}%

\usepackage{tikz}
\usetikzlibrary{shapes,arrows,positioning,fit,backgrounds,calc,shadows}
\usepackage{enumitem}
\usepackage{hyperref} 
\usepackage{bm} 

\usepackage{pdfpages} 
\usepackage{mdframed}

\usepackage[capitalise]{cleveref}
\usepackage[activate={true,nocompatibility},final,kerning=true,factor=1100,stretch=10,shrink=10]{microtype}
\usepackage{academicons}
\usepackage{fontawesome5}
\definecolor{orcidlogocol}{rgb}{0.65, 0.807, 0.223}
\newcommand{\orcidam}[1]{$\,$\href{https://orcid.org/#1}{\textcolor{orcidlogocol}{\faOrcid}}}

\theoremstyle{thmstyleone}%
\theoremstyle{thmstyletwo}%

\theoremstyle{thmstylethree}%

\begin{document}

\title[AI Cosmologist I: Agentic Data Analysis]{The AI Cosmologist I: An Agentic System for Automated Data Analysis}


\author*[1]{\fnm{Adam} \sur{Moss}\orcidam{0000-0002-7245-7670}}\email{adam.moss@nottingham.ac.uk}

\affil*[1]{\orgdiv{School of Physics and Astronomy}, \orgname{University of Nottingham}, \orgaddress{\street{University Park}, \city{Nottingham}, \postcode{NG7 2RD}, \country{UK}}}


\abstract{We present the AI Cosmologist, an agentic system designed to automate cosmological/astronomical data analysis and machine learning research workflows. This implements a complete pipeline from idea generation to experimental evaluation and research dissemination, mimicking the scientific process typically performed by human researchers. The system employs specialized agents for planning, coding, execution, analysis, and synthesis that work together to develop novel approaches. Unlike traditional auto machine-learning systems, the AI Cosmologist generates diverse implementation strategies, writes complete code, handles execution errors, analyzes results, and synthesizes new approaches based on experimental outcomes. We demonstrate the AI Cosmologist capabilities across several machine learning tasks, showing how it can successfully explore solution spaces, iterate based on experimental results, and combine successful elements from different approaches. Our results indicate that agentic systems can automate portions of the research process, potentially accelerating scientific discovery. The code and experimental data used in this paper are available on GitHub at \url{https://github.com/adammoss/aicosmologist}. Example papers included in the appendix demonstrate the system's capability to autonomously produce complete scientific publications, starting from only the dataset and task description.}


\keywords{cosmology, artificial intelligence, machine learning, automated research, agentic systems, large language models}

\maketitle


\section{Introduction}\label{sec:intro}

Cosmology and astrophysics are entering a fundamentally data-rich era. Upcoming surveys and experiments, such as the Vera C.~Rubin Observatory \cite{LSST:2008ijt}, the \textit{Euclid} space telescope \cite{EUCLID:2011zbd}, the Square Kilometre Array (SKA) \cite{SKA:2018ckk}, and spectroscopic surveys like DESI \cite{DESI:2016fyo}, will produce unprecedented volumes of high-dimensional data. Cosmological simulations further contribute to this data abundance by generating thousands of high-resolution simulations (e.g \cite{Villaescusa-Navarro:2019bje}). Such large-scale data sets, characterized by petabyte-to-exabyte scales, render traditional manual or semi-manual analysis workflows impractical \cite{2019arXiv190407248B}. Efficient extraction of scientific insights thus requires transformative approaches to data analysis.

Machine learning (ML) has become essential in astrophysics, successfully automating tasks such as galaxy classification \cite{2015MNRAS.450.1441D}, supernova detection \cite{2017ApJ...837L..28C} transient detection \cite{2012PASP..124.1175B}, strong lens discovery \cite{2017MNRAS.472.1129P}, photometric redshift estimation \cite{2004PASP..116..345C}, and gravitational waves \cite{2018PhRvD..97d4039G}. Beyond astronomy, ML is increasingly critical across sciences, achieving breakthroughs in fields like materials discovery \cite{Butler2018MLMaterials}, protein folding prediction \cite{Jumper2021AlphaFold}, and Earth system modeling \cite{Reichstein2019ESDLearning}. These successes underscore ML's capability to manage complexity and scale where traditional techniques falter.

Yet, practical ML deployment still demands significant domain-specific expertise and iterative experimentation, creating bottlenecks in research workflows. Automated Machine Learning (AutoML) systems, designed to minimize human intervention, have emerged to address this challenge \cite{zoller2021benchmark}. While AutoML can streamline model selection, hyperparameter tuning, and feature engineering, existing solutions typically optimize predefined workflows and struggle with novel tasks requiring creative problem-solving or iterative refinement.

Advances in Large Language Models (LLMs), exemplified by OpenAI Codex \cite{Chen2021Codex} and AlphaCode \cite{Li2022AlphaCode}, offer complementary opportunities for workflow automation. These models excel at generating executable code and facilitating human-like reasoning, enabling high-level automation previously unattainable. Building on these innovations, agentic frameworks that embed LLM-based reasoning into autonomous decision-making loops are emerging as transformative tools. Frameworks such as ReAct \cite{Yao2023ReAct}, Reflexion \cite{Shinn2023Reflexion}, and domain-specific agents like SWE-agent \cite{yang2024swe} and ChemCrow \cite{m2024augmenting} demonstrate significant potential in automating complex, iterative scientific and engineering tasks.

In this paper, we introduce the \textbf{AI Cosmologist}, an agentic system designed to automate end-to-end data analysis in cosmology. Our framework integrates AutoML techniques, LLM-driven code generation, and autonomous reasoning agents to facilitate fully automated scientific workflows. The AI Cosmologist autonomously formulates hypotheses, designs computational experiments, evaluates results, and iteratively refines methods without manual intervention.

The structure of the paper is as follows. Section 2 reviews related work in automated machine learning, AI-assisted programming, and agentic systems. Section 3 details our agentic system architecture, explaining the specialized agents, their coordination, and both the research and dissemination workflows. Section 4 presents experimental results on two representative cosmological machine learning tasks: galaxy morphology classification and cosmological parameter inference. Section 5 discusses the implications of our findings, current limitations, and promising future directions. Section 6 concludes with a summary and discussion. The appendix includes two complete research papers autonomously generated by the AI Cosmologist system.

\subsection{Contributions}

Specifically, our contributions include:
\begin{itemize}
\item A novel agentic system for automating scientific ML workflows in cosmology and astronomy, combining AutoML and LLM-driven automation with iterative reasoning;
\item Integration of LLM-based agents for creative and dynamic research pipeline construction and execution;
\item Demonstration of state-of-the-art performance on representative tasks, including galaxy morphology classification and cosmological parameter inference;
\item A comprehensive, autonomous research pipeline capable of producing publication-ready results and visualizations;
\item Empirical validation through high quality scientific papers presented in the appendix.
\end{itemize}


\section{Related Work}\label{sec:related}

\subsection{Automated Machine Learning (AutoML)}\label{subsec:automl}
AutoML automates pipeline design, model selection, and hyperparameter tuning \cite{Hutter2019,Elsken2019,zoller2021benchmark}. Early frameworks like Auto-WEKA \cite{Thornton2013} and auto-sklearn \cite{Feurer2015} combined Bayesian optimization with meta-learning \cite{Vanschoren2019} to efficiently explore ML pipelines. Evolutionary approaches, notably TPOT \cite{Olson2016}, further automate pipeline optimization via genetic algorithms.

Neural architecture search (NAS) extended AutoML to deep learning, achieving human-competitive results through reinforcement learning \cite{Zoph2017}, evolutionary strategies \cite{Real2019}, and differentiable methods like DARTS \cite{Liu2019}. Meta-learning approaches, such as MAML \cite{Finn2017}, facilitate rapid adaptation across tasks, further enhancing AutoML's efficiency and generality.

AutoML has shown significant promise in scientific domains. Applications include automated detection of asteroids in Hubble imagery \cite{Kruk2022} and morphology classification of galaxies \cite{Tarsitano2022}. However, these systems typically explore predefined search spaces and still require substantial human guidance in handling complex scientific problems.

\subsection{AI-Assisted Programming}\label{subsec:ai_prog}
AI-assisted programming has rapidly evolved, driven by large language models (LLMs) trained on code. OpenAI's Codex \cite{Chen2021} significantly advanced the field by achieving strong performance on benchmarks like HumanEval, demonstrating AI's ability to translate natural language into executable code.

Following Codex, sophisticated models such as CodeGen \cite{Nijkamp2022}, AlphaCode \cite{Li2022AlphaCode}, InCoder \cite{Fried2023}, and StarCoder \cite{Li2023StarCoder} have emerged, each introducing innovations like multi-turn synthesis, code infilling, and extensive contextual understanding. These models offer potential for substantial productivity gains, error reduction, and improved code quality, especially in complex scientific codebases.

In scientific computing contexts, AI-assisted programming can streamline tasks such as rapid prototyping of ML models, data pipeline standardization, and error detection. Domain-specific coding assistants are becoming increasingly feasible, promising tailored AI support that understands specialized scientific languages and workflows.

\subsection{Agentic Systems and Autonomous Scientists}\label{subsec:agentic}
Agentic systems embed reasoning and action capabilities into AI, enabling autonomous planning, execution, and iterative improvement. Frameworks such as ReAct \cite{Yao2023}, Reflexion \cite{Shinn2023}, and HuggingGPT \cite{Shen2023} augment LLMs with tool use, reflection mechanisms, and external memory to create adaptive problem-solving agents.

Recent advances include autonomous domain-specific systems like ChemCrow for chemistry \cite{m2024augmenting}, SWE-agent for software engineering \cite{yang2024swe}, and autonomous laboratory systems like ChemGPT \cite{Boiko2023}. These agents actively manage research cycles, from hypothesis formation and experimental design to iterative refinement based on empirical outcomes.

In astrophysics and cosmology, early explorations have demonstrated potential for agentic systems in simulation-based inference and automated scientific discovery \cite{Cranmer2020,Brehmer2019,Xu2023ExpertIter,Laverick:2024fyh}. AI Cosmologist builds upon these foundations, uniquely focusing on automating the full scientific ML lifecycle in cosmology, integrating interpretability and performance critical to scientific understanding.


\section{An Agentic System for Automated Cosmological Data Analysis}\label{sec:system}

\subsection{Agent Architecture}\label{subsec:arch}

The AI Cosmologist employs a modular architecture consisting of specialized components for different stages of the machine learning research process. This design follows the principles of agentic systems where autonomous software components (agents) are designed to perform specific functions toward a common goal while maintaining their independent decision-making capabilities.
At the core of this system are Large Language Models (LLMs)—neural network architectures trained on vast text corpora that can generate contextually relevant text based on inputs. These models function by predicting probable token sequences, implementing sophisticated attention mechanisms that allow them to maintain context coherence over extended interactions. In our system, each agent leverages an LLM configured with specialized instructions that define its domain of operation, constraints, and objectives:
\begin{itemize}
\item \textbf{Planning Agent:} Generates implementation plans and strategies based on task specifications and dataset details. This agent employs prompt engineering techniques to optimize for scientific reasoning and hypothesis generation.
\item \textbf{Coding Agent:} Converts plans into executable ML code. Specialized with code-specific instructions, this agent leverages the LLM's understanding of programming patterns and scientific computing libraries to produce functionally correct implementations.

\item \textbf{Execution Agent:} Runs the generated code and handles errors. This agent combines an LLM with external tool integration, enabling the execution of code in controlled environments and the interpretation of runtime outputs, including errors and performance metrics.

\item \textbf{Analysis Agent:} Evaluates results and generates insights. This agent processes experimental outputs, applying statistical reasoning to interpret model performance and identify strengths and weaknesses in the implemented approach.

\item \textbf{Synthesis Agent:} Creates new approaches by combining successful elements from previous runs. Implementing a meta-learning capability, this agent reasons across multiple experiments to identify patterns and generate novel approaches.

\item \textbf{Literature Agent:} Connects research implementations to the scientific literature by automatically querying repositories such as arXiv and INSPIRE-HEP. This agent retrieves relevant papers, extracts their content, and identifies methodological approaches and benchmark results.  The agent maintains a structured bibliography of relevant papers with their citations, summaries, and relevance assessments.

\end{itemize}

The coordination between these agents follows a directed graph structure, where the output of one agent serves as input to another. This orchestration is managed through a conditional execution framework where subsequent agent invocations depend on the success and content of previous operations. Each agent maintains its own context window containing relevant information for its specific task, while a global context preserves key information across the entire system.

Each agent can access external tools when necessary, including code execution environments, data visualization libraries, and scientific computing frameworks. These tool integrations extend the system's capabilities beyond text generation, enabling it to interact with computational resources and perform concrete operations on data.

To illustrate how these agents function in practice, Fig.~\ref{fig:planning-prompt} gives the actual prompt used by the Planning Agent to generate initial ideas. This exemplifies how the system structures LLM interactions to elicit specific types of scientific reasoning. The agent is instructed to adopt an expert scientist persona, provided with task-specific context, and guided with precise formatting requirements.

\begin{figure}
\begin{mdframed}
\begin{verbatim}
You are an expert scientist. Your task is to come up
with a set of {num_ideas} implementation plans to
perform the task given the additional information
below.

* Task *

{task}

* Dataset Information *

{additional_info}

* Instructions *

- Generate {num_ideas} plans suitable for
implementation  in PyTorch or some other ML framework.
- These should be high level plans, giving high-level details of 
the model and training procedure. It is not necessary 
to give the exact details of the model architecture or training 
procedure.
- Each plan should have new, different aspects
compared to other plans.
- It is fine to reuse ideas, but each should have
some originality.
- Each plan should include techniques that logically
integrate.
- Do not try to do too much at once.
- Ensure each plan is scientifically sound.
- Think deeply about the scientific motivation for the
plan, justifying each plan against the task and data.
- Do not write any code yet.
- Present each plan in an idea code block
\end{verbatim}
\end{mdframed}
\caption{Example prompt template used by the Planning Agent to generate initial implementation ideas. Curly braces indicate variable placeholders that are dynamically filled with task-specific information.}
\label{fig:planning-prompt}
\end{figure}

The agent operates through a structured workflow that encompasses two distinct phases: (1) the research phase and (2) the dissemination phase.

\subsection{Research Phase}\label{subsec:research_wf}

\begin{figure}
    \centering
    \includegraphics[width=\columnwidth]{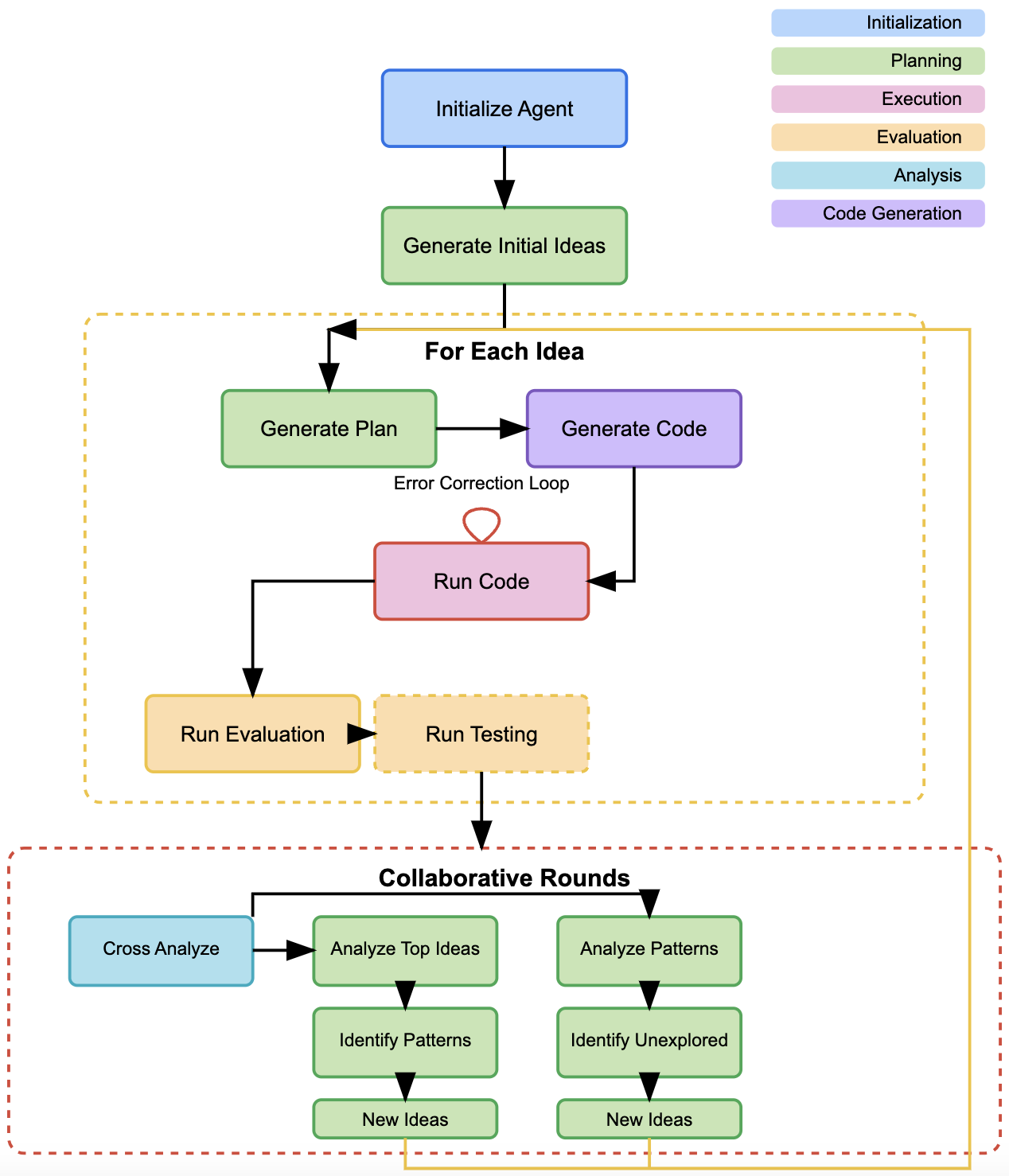} 
    \caption{Workflow diagram of the AI Cosmologist system in the research phase. The process begins with initialization and generation of initial ideas, followed by a development cycle for each idea that includes planning, code generation, execution, and evaluation. The system then enters collaborative rounds where cross-analysis of results leads to two parallel pathways: analyzing top-performing ideas to identify successful patterns, and examining the solution space to discover unexplored approaches. }
    \label{fig:agent-flow}
\end{figure}

 The research phase implements a complete scientific discovery cycle, systematically generating, testing, and refining hypotheses through experimentation, as illustrated in Fig.~\ref{fig:agent-flow}. This phase begins with initialization and idea generation, proceeds through implementation and evaluation for each idea, and culminates in collaborative rounds that synthesize insights across multiple experiments to generate improved approaches.

\subsubsection{Initialization}\label{subsubsec:init}

The system begins by loading dataset information from files that describe the problem and data characteristics, $D$. Rather than requiring exhaustive details, the system works with high-level information about the dataset—such as file structure, data types, and column names for tabular data. This information may include relevant scientific background to contextualize the problem. Additionally, the system requires a task specification, $T$, which can be as straightforward as "minimize the MSE loss on test data" or more specific research objectives.

\subsubsection{Idea Generation}\label{subsubsec:idea_gen}
The planning agent generates multiple distinct implementation approaches:

\begin{equation}
I = \{i_1, i_2, ..., i_n\}\,,
\end{equation}
where each idea $i_j$ represents a unique strategy for addressing the task. These initial ideas are stored in a centralized repository for tracking throughout the research process.

\subsubsection{Plan Development}\label{subsubsec:plan_dev}
For each idea $i_j$, the agent develops a comprehensive implementation plan $P_j$:

\begin{equation}
P_j = f_\text{plan}(i_j, D, T)\,,
\end{equation}
Plans detail all aspects of implementation including data loading, preprocessing, model architecture, training procedures, and evaluation methods.

The agent can perform multiple reflection steps to refine plans:

\begin{equation}
P_j^{(k+1)} = f_\text{reflect}(P_j^{(k)}, D, T)\,,
\end{equation}
where $P_j^{(k)}$ represents the plan at reflection step $k$.

\subsubsection{Code Implementation}\label{subsubsec:code_impl}
The coding agent transforms plans into executable code through a systematic process represented by the following equation:
\begin{equation}
C_j = f_{\text{code}}(P_j, D, T)
\end{equation}
The generated code includes complete machine learning implementations that cover all requirements for end-to-end execution. These implementations feature data loading and preprocessing pipelines designed to appropriately handle the transformation and augmentation of input data. Additionally, the agent incorporates training and evaluation procedures with appropriate optimization methods, loss functions, and metrics tailored to the specific task requirements. The code integrates with experiment tracking tools to systematically log metrics, hyperparameters, and visualizations throughout the training process. Further functionality includes checkpoint handling mechanisms that enable saving model states and resuming training when necessary. Finally, the implementation provides result visualization capabilities to generate informative plots and visual representations that facilitate interpretation of model performance.

Self-reflection mechanisms enable code refinement through an iterative process:
\begin{equation}
C_j^{(k+1)} = f_\text{code\_reflect}(C_j^{(k)}, P_j, D, T)
\end{equation}
where $C_j^{(k)}$ represents the code at reflection step $k$.

\subsubsection{Execution and Evaluation}\label{subsubsec:exec_eval}
The execution agent runs the generated code on the target dataset:
\begin{equation}
R_j = f_{\text{execute}}(C_j, D)
\end{equation}
where $R_j$ represents the results of executing code $C_j$. When errors occur during execution, the agent diagnoses the issues and generates code fixes using a diff-based editing format implemented through the open source Aider package\footnote{\url{https://github.com/Aider-AI/aider}}. This approach saves significant LLM tokens by only outputting the specific changes to the codebase rather than regenerating the entire implementation for each fix:
\begin{equation}
C_j^{(\text{fixed})} = f_{\text{error\_fix}}(C_j, E)
\end{equation}
where $E$ represents detected errors. The diff-based editing allows for precise modifications to address specific issues while maintaining the broader code context. This process continues until successful execution or until reaching the maximum number of retry attempts. If code errors cannot be resolved, the agent will mark the approach as unsuccessful. 

\subsubsection{Synthesis}\label{subsubsec:synthesis}
The synthesis agent performs comprehensive cross-idea analysis to evaluate and compare the effectiveness of different implementation approaches. This process begins with a ranking function that assesses all experimental results:
\begin{equation}
\text{Rank} = f_{\text{rank}}(\{R_1, R_2, \ldots, R_n\})
\end{equation}
The agent then follows two parallel pathways to generate new ideas. The first pathway analyzes top-performing ideas to identify successful patterns:
\begin{equation}
\text{Patterns} = f_{\text{patterns}}(\text{Rank}, \{R_1, R_2, \ldots, R_n\}, \{P_1, P_2, \ldots, P_n\})
\end{equation}
A second analysis pathway examines the entire solution space to identify unexplored regions:
\begin{equation}
\text{Unexplored} = f_{\text{unexp}}(\{R_1, R_2, \ldots, R_n\}, \{P_1, P_2, \ldots, P_n\})
\end{equation}
These complementary analyses facilitate the generation of two types of idea. First, new, iterative ideas
\begin{equation}
I^{(\text{iter})} = f_{\text{iter}}(\text{Patterns}, \text{Rank}, \{R_1, \ldots, R_n\}, \{P_1, \ldots, P_n\})
\end{equation}
and second, new diverse ideas
\begin{equation}
I^{(\text{diverse})} = f_{\text{diverse}}(\text{Unexplored}, \{R_1, \ldots, R_n\}, \{P_1, \ldots, P_n\})
\end{equation}

The final set of new ideas combines both synthesized improvements based on successful approaches and novel ideas that explore previously unexamined regions of the solution space.

\subsubsection{Collaborative Rounds}\label{subsubsec:collab}
Following the cross-analysis phase, the newly generated ideas are fed back into the planning stage to initiate additional cycles of development. This cyclical process continues for a predetermined number of collaborative rounds, with each round building upon insights gained from previous iterations. Each collaborative round processes the new ideas through the full pipeline of planning, coding, execution, and evaluation, enabling progressive refinement of approaches based on accumulated experimental evidence.

\subsection{Research Dissemination Phase}\label{subsec:dissemination}

The dissemination phase activates after promising results are obtained from the research phase, focusing on transforming experimental outcomes into comprehensive scientific communications. This phase employs multiple specialized components working in concert to produce publication-ready materials.

At the core of this process is a structured workflow that begins with the systematic planning of the scientific narrative. The Planning Agent first evaluates experimental results to identify key findings, contributions, and their significance within the broader scientific context. This analysis generates a detailed paper outline including proposed titles, section structures, key results to highlight, and necessary literature connections.

The Literature Agent then conducts comprehensive searches across scientific repositories such as arXiv and INSPIRE-HEP to retrieve relevant publications. This agent employs carefully crafted queries to identify papers related to the methodology, dataset, and research domain. For each retrieved paper, the agent extracts metadata (authors, citations, publication venue), performs content analysis to identify relevant methods and results, and creates structured summaries with relevance assessments. These literature connections enable proper attribution of methods and positioning of results relative to the current state of the art.

Building on the paper plan and literature review, the system generates complete section drafts following conventional scientific publication structure (abstract, introduction, related work, methodology, results, discussion, conclusion). Each section is crafted with appropriate technical depth, mathematical precision, and visual elements. For methodology sections, the agent extracts implementation details from experimental code while translating algorithmic components into precise mathematical notation. Results sections incorporate automatically generated visualizations of experimental outcomes, including comparison plots with baseline methods and state-of-the-art results identified in the literature.

The final output of the dissemination phase includes:
\begin{itemize}
    \item A complete scientific manuscript in LaTeX format with appropriate sectioning, citations, and mathematical notation
    \item High-quality visualizations of experimental results in publication-ready formats
    \item A comprehensive bibliography in BibTeX format with entries for all referenced works
    \item Compiled PDF documents ready for review or submission
\end{itemize}

This automated research dissemination capability represents a significant advancement in scientific AI systems, enabling the full research cycle from idea generation through experimentation to publication-ready communication without manual intervention. However, it is important to note that the current system generates drafts that benefit from human review and refinement before formal submission to scientific venues.

\subsection{Implementation Details}\label{subsec:impl_details}

The AI Cosmologist system is implemented using a carefully selected combination of state-of-the-art technologies that balance performance requirements with practical considerations. Large language models serve as the foundation for all agent components within the system architecture. We employ Gemini 2.5 Pro (API version \texttt{gemini-2.5-pro-exp-03-25}) for the majority of agent functions, including planning, analysis, and synthesis tasks. This model was selected due to its current top-ranking performance on science-based reasoning and code development benchmarks. Gemini 2.5 Pro offers considerable advantages for our implementation, including free usage up to a reasonable rate limit of 50 requests per day, which facilitates both development and limited-scale experimental deployments.

For the diff-based code editing components, we implement a different approach using OpenAI's \texttt{o3mini-high} model. This specific choice was made because code editing often requires more frequent API requests within a single experimental cycle, making the rate-limited Gemini model potentially restrictive for this particular task. The o3mini model carries a defined cost structure of around \$1 per million tokens for input, and \$5 per million tokens for output. Despite these costs, we find this model provides efficient performance for the code editing task while maintaining reasonable expenses. Our empirical measurements indicate that each complete end-to-end experimental process typically costs several dollars, representing an acceptable expense given the computational complexity and potential scientific value of the automated research performed.


\section{Experimental Results}\label{sec:results}

\subsection{Experimental Setup}\label{subsec:setup}

We evaluated the AI Cosmologist system on two representative cosmological machine learning tasks to demonstrate its capabilities. While future work will include more exhaustive studies across a wider range of problems, the current experiments serve as a proof of concept, illustrating the methodology's efficacy and potential. The two datasets chosen represent distinct challenges in cosmological analysis: galaxy morphology classification and cosmological parameter inference.

The first dataset is derived from the Galaxy Zoo 2 (GZ2) project \cite{Willett:2013wda}, which provides detailed morphological classifications for 304,122 galaxies from the Sloan Digital Sky Survey (SDSS). This dataset presents a challenging regression task where the objective is to predict 37 morphological probability values for each galaxy image based on the GZ2 decision tree. The dataset was made available through a Kaggle competition\footnote{\url{https://www.kaggle.com/competitions/galaxy-zoo-the-galaxy-challenge/}}, providing a standardized evaluation framework with clear performance metrics. The task encompasses several computer vision challenges including feature extraction from noisy astronomical images, handling of orientation and scale variance, and modeling the probabilistic nature of human classifications.

The second dataset utilizes the Quijote simulation suite \cite{Villaescusa-Navarro:2019bje}, specifically designed for cosmological parameter inference tasks. We worked with the Latin Hypercube subset comprising 2000 simulations that systematically explore a five-dimensional parameter space: the matter density parameter $\Omega_\mathrm{m}$, the baryon density parameter $\Omega_\mathrm{b}$, the dimensionless Hubble parameter $h$, the primordial spectral index $n_\mathrm{s}$, and the amplitude of matter fluctuations $\sigma_8$. Each simulation provides a dark matter density field discretized on a $64^3$ grid within a cubic volume of $(1 \, \mathrm{Gpc}/h)^3$, representing the spatial distribution of dark matter at redshift $z=0$. This dataset presents a complex regression task requiring the model to extract subtle features from 3D density fields that correlate with fundamental cosmological parameters.

For each dataset, the AI Cosmologist was provided with only basic information about the data structure and the task objective. The system was then allowed to autonomously generate multiple implementation strategies, evaluate their performance, and iteratively refine its approaches through collaborative rounds. We tracked the progression of model performance across iterations to evaluate both the absolute quality of solutions and the system's ability to improve through iterative refinement.

To ensure a comprehensive exploration of the solution space while maintaining computational efficiency, we configured the AI Cosmologist with the following hyperparameters. The system initially generated 20 distinct implementation ideas for each task, providing a diverse foundation of approaches. Each idea underwent complete development through the planning, coding, execution, and evaluation phases. Following this initial exploration, we conducted 5 collaborative rounds, with each round generating 6 new ideas (3 based on synthesis of top-performing approaches and 3 exploring novel directions). This resulted in a total of 50 implementation attempts per dataset, providing sufficient coverage to demonstrate the system's ability to progressively refine solutions while exploring the solution space. For error correction, we allowed a maximum of 3 retry attempts for each implementation to resolve runtime issues before considering an approach unsuccessful.

\subsection{Galaxy Zoo Results}\label{subsec:gz_results}

For the Galaxy Zoo experiment, the AI Cosmologist was tasked with a straightforward objective:

\begin{mdframed}
\begin{verbatim}
Obtain the minimum MSE error on test data.
Ensure to output RMSE as part of the evaluation.
\end{verbatim}
\end{mdframed}

Following each successful implementation, the agent automatically submitted predictions to the Kaggle competition platform via its API and recorded the public leaderboard score. These scores, along with additional evaluation metrics collected during training and validation, provided quantitative feedback that informed subsequent refinement cycles.

Figure~\ref{fig:idea_progression} illustrates the evolution of implementation strategies across the three main phases of the research workflow. In the initial ideation phase, the agent generated a diverse set of approaches primarily centered around fine-tuning pre-trained convolutional neural networks with various architectures (ResNet-50, EfficientNet-B4, Vision Transformer) and training configurations. These initial ideas explored different data augmentation techniques, optimization strategies, and model architectures while maintaining a common thread of addressing the regression nature of the morphological classification task.

\begin{figure*} 
\centering
\begin{tikzpicture}[
    node distance=1.5cm,
    box/.style={
        rectangle,
        rounded corners,
        draw=black,
        thick,
        fill=blue!10,
        text width=15cm, 
        minimum height=3cm,
        align=left,
    },
    idea/.style={
        rectangle,
        rounded corners,
        draw=black,
        thick,
        fill=green!10,
        text width=15cm, 
        minimum height=4cm,
        align=left,
    },
    annotation/.style={
        rectangle,
        draw=none,
        fill=none,
        text width=5cm,
        align=center,
        font=\bfseries\Large
    },
    arrow/.style={
        ->,
        >=stealth,
        thick,
        draw=black
    }
]


\node[idea] (ideas) {
    \textbf{\LARGE Initial Ideas}\\[0.3cm]
    \begin{minipage}{0.98\linewidth} 
    \begin{itemize}[leftmargin=*, itemsep=0.1cm] 
        \item Fine-tune pre-trained ResNet-50 CNN with image augmentations for galaxy morphology regression using MSE loss, AdamW optimizer and ReduceLROnPlateau learning rate scheduler.
        \item Train a fine-tuned pre-trained EfficientNet-B4 CNN (384px images) with custom normalization, data augmentation, AdamW optimizer, cosine annealing learning rate scheduler and MSE loss.
        \item Fine-tune pre-trained Vision Transformer with data augmentation, AdamW optimizer and learning rate scheduler, replacing the final classification layer for regression, minimizing MSE loss, and clamping predictions.
        \item Fine-tune pre-trained ResNet-50 with random rotation, flips, and color jitter, using MSE loss, AdamW optimizer, and ReduceLROnPlateau learning rate scheduler.
        \item Ensemble N independently trained pre-trained ResNet-50 models with data augmentation, AdamW optimizer, ReduceLROnPlateau scheduler, and MSE loss, averaging clamped predictions for the Galaxy Zoo challenge.
        \item \ldots 
    \end{itemize}
    \end{minipage}
};

\draw[arrow, line width=1.5mm] (ideas.south) -- ++(0,-1.2) node[midway, right] {\textbf{Evaluation \& Synthesis}};


\node[box, below=1.1cm of ideas] (analysis) {
    \textbf{\LARGE Analysis Stage}\\[0.3cm]
    \begin{minipage}{0.48\linewidth}
    \textbf{STRENGTHS:}
    \begin{itemize}[leftmargin=*, itemsep=0.1cm]
        \item Pre-trained CNNs (EfficientNet, ConvNeXt, ResNet) are crucial.
    \end{itemize}
    \textbf{WEAKNESSES :}
    \begin{itemize}[leftmargin=*, itemsep=0.1cm]
        \item Training loss significantly lower than validation loss/RMSE.
        \item MSE treats all classes equally.
        \item Advanced augmentations, fine-tuning strategies, target dependency modeling are lacking.
    \end{itemize}
    \end{minipage} \hfill 
    \begin{minipage}{0.48\linewidth}
    \textbf{OPPORTUNITIES:}
    \begin{itemize}[leftmargin=*, itemsep=0.1cm]
        \item Advanced Augmentation Strategies: Mixup/CutMix, domain-specific augmentations, TTA.
        \item Weighted MSE, hierarchical loss, target transformation.
    \end{itemize}
    \textbf{DIVERSITY:}
    \begin{itemize}[leftmargin=*, itemsep=0.1cm]
        \item Self-Supervised Learning (SSL).
        \item Explicit Hierarchical Modeling.
    \end{itemize}
     \ldots 
    \end{minipage}
};

\draw[arrow, line width=1.5mm] (analysis.south) -- ++(0,-1.2) node[midway, right] {\textbf{Refinement \& Integration}};


\node[idea, below=1.2cm of analysis, fill=yellow!10] (new_ideas) {
    \textbf{\LARGE New Ideas}\\[0.3cm]
     \begin{minipage}{0.98\linewidth} 
    \begin{itemize}[leftmargin=*, itemsep=0.1cm] 
        \item Fine-tune pre-trained EfficientNet-B4 CNN with hierarchical MSE loss, domain-specific augmentations (PSF simulation, noise injection), and AdamW optimizer.
        \item Fine-tune pre-trained Vision Transformer with regression-adapted Mixup/CutMix, ImageNet normalization, AdamW optimizer, cosine annealing scheduler and MSE loss.
        \item Ensemble fine-tuned EfficientNet-B3/B4, ConvNeXt-Tiny models with multi-scale inputs, ImageNet pre-training, AdamW, ReduceLROnPlateau/CosineAnnealingLR, MSE loss with logit target transformation and HPO.
        \item Train a Vision Transformer (ViT) using self-supervised learning (SSL) on combined galaxy training and test images, then fine-tune it with a regression head and MSE loss.
        \item CNN/Vision Transformer with hierarchical loss and conditional output heads to explicitly model the Galaxy Zoo decision tree's conditional probability structure.
        \item  \ldots 
    \end{itemize}
    \end{minipage}
};
\end{tikzpicture}
\caption{Evolution of implementation strategies for the Galaxy Zoo dataset. The figure shows a subset of the initial ideas, analysis of experimental results, and new, synthesized ideas. Text has been abbreviated and annotated for space considerations.}
\label{fig:idea_progression}
\end{figure*}

The analysis phase revealed several key insights that guided subsequent refinements. The agent identified that pre-trained CNNs, particularly newer architectures like EfficientNet and ConvNeXt, consistently outperformed other approaches. However, it also noted significant weaknesses, including substantial gaps between training and validation performance (suggesting overfitting) and suboptimal handling of class dependencies inherent in the Galaxy Zoo decision tree structure. The analysis highlighted opportunities for improvement through advanced augmentation strategies (such as Mixup/CutMix and domain-specific transformations), more sophisticated loss functions that account for the hierarchical nature of the classification task, and target transformations that better capture the probability distribution characteristics.

In the refinement phase, the agent synthesized these insights to generate more sophisticated implementations. These new approaches incorporated hierarchical loss functions to model the Galaxy Zoo decision tree dependencies, employed domain-specific augmentations tailored to astronomical imaging, and explored self-supervised learning to leverage unlabeled data. Particularly notable was the development of multi-resolution input strategies and ensemble approaches that combined complementary model architectures.

\begin{figure}[h] 
    \centering
    \includegraphics[width=\columnwidth]{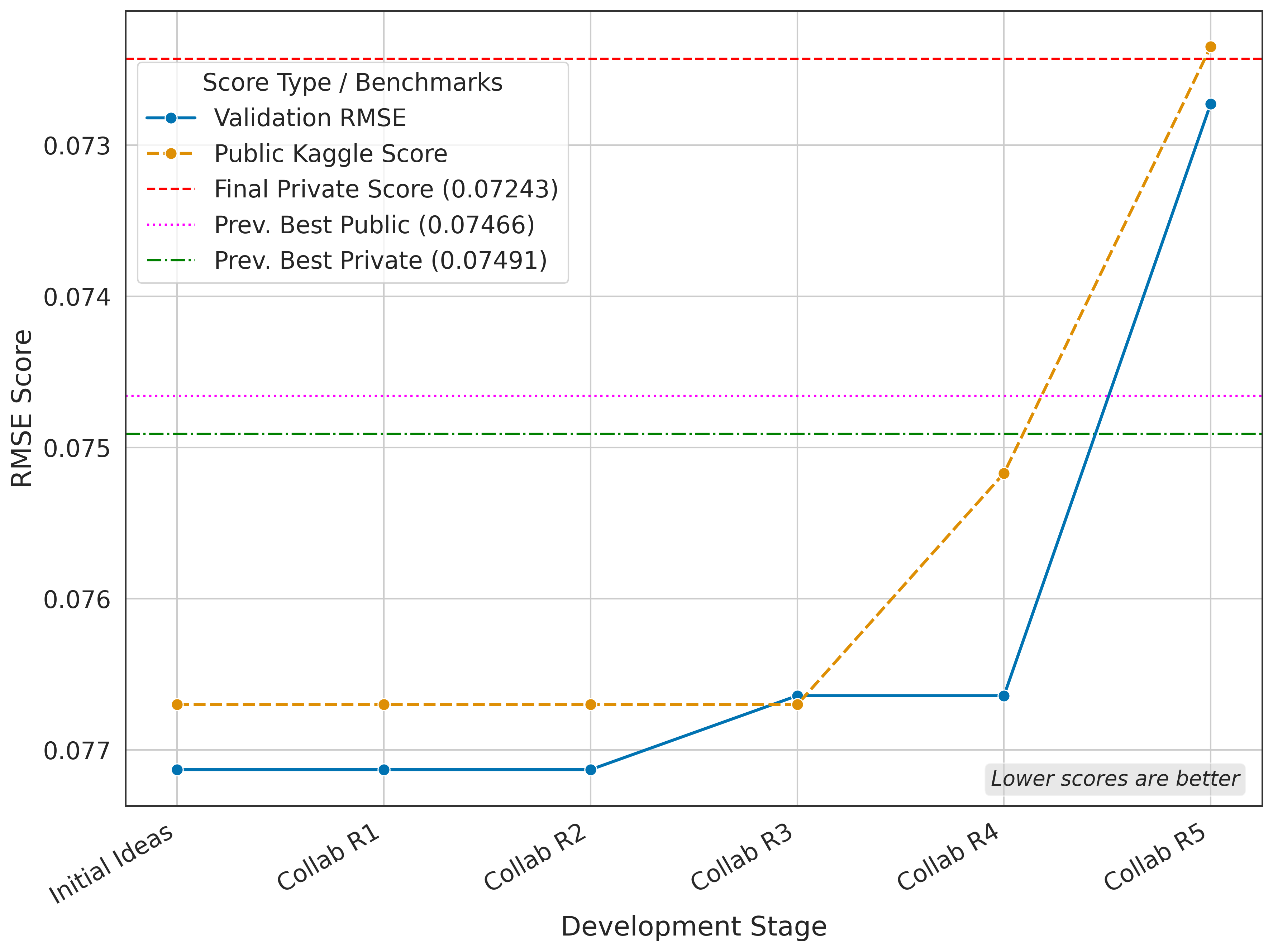} 
    \caption{Improvement of the best validation RMSE and public Kaggle score on the Galaxy Zoo 2 dataset, through initial ideas to collaborative rounds. }
    \label{fig:result_progression}
\end{figure}

Figure~\ref{fig:result_progression} tracks the performance improvement across successive implementation cycles. The initial ideas achieved respectable but not exceptional performance, with RMSE values around 0.077 on the validation set and corresponding Kaggle public scores. Through iterative refinement, performance steadily improved, with particularly significant gains observed in the final collaborative round. The breakthrough implementation combined test-time augmentation with multi-scale resizing—a strategy that substantially improved robustness to orientation and scale variations in galaxy images. This approach achieved an RMSE of 0.07235 on the Kaggle public leaderboard, representing a substantial improvement over earlier implementations.

Most notably, when evaluated on the private Kaggle leaderboard (which was not accessible during development), this final implementation exceeded the performance of the original competition winner. This demonstrates the AI Cosmologist's ability to not only autonomously develop effective solutions but to discover novel implementation strategies that match or exceed human expert performance.

The complete research cycle culminated in the automated generation of a scientific paper detailing the methodology and results, included in the appendix. This paper was produced entirely by the system without human intervention.

\subsection{Quijote Results}\label{subsec:quijote_results}

For the Quijote Results experiment, the AI Cosmologist was tasked with the objective:

\begin{mdframed}
\begin{verbatim}
Obtain the minimum MSE error on test data.
Ensure to output MSE, MAE and R2, for each parameter as part of 
the evaluation.
\end{verbatim}
\end{mdframed}

The initial ideation phase produced diverse approaches primarily based on 3D convolutional neural networks with various enhancements such as ResNet-style residual connections, Inception-inspired blocks, and attention mechanisms. These implementations consistently employed log-transformation of density fields and standardization of both inputs and target parameters. Physical symmetries were respected through data augmentations like random rotations and flips. Several variations emerged, from standard 3D CNNs to more sophisticated Vision Transformers, contrastive pre-training methods, and dual-branch networks processing both spatial and spectral information.

Analysis of these initial implementations revealed important patterns. The system identified the effectiveness of 3D CNNs for spatial feature extraction and the necessity of proper preprocessing strategies. However, it also recognized significant challenges, particularly the difficulty in accurately predicting certain parameters ($\Omega_b$ and $h$) compared to others ($\Omega_m$ and $\sigma_8$), computational constraints in 3D model training, and information loss in standard pooling operations. These insights guided the identification of key opportunity areas focusing on physics-informed architectures, uncertainty quantification, and multi-scale feature extraction.

In the refinement phase, the AI Cosmologist developed more sophisticated approaches that substantially advanced beyond initial implementations. These newer models incorporated multivariate probability distribution modeling, multi-scale feature extraction with auxiliary tasks, self-supervised pre-training, and physics-informed feature representations. Performance evaluation demonstrated clear progression across iterations, with the most significant improvements coming from approaches that explicitly incorporated physical insights into the model architecture.

The most successful implementation was a physics-augmented 3D CNN that combined deep feature extraction with explicitly computed power spectrum and density probability distribution features. This hybrid approach achieved state-of-the-art performance by significantly improving constraints on the traditionally challenging $\Omega_b$ and $h$ parameters while maintaining excellent accuracy for $\Omega_m$ and $\sigma_8$. The research culminated in an automatically generated scientific paper included in the appendix.

\section{Discussion}\label{sec:discussion}

Our experiments demonstrate that the AI Cosmologist system can successfully generate diverse, executable implementations for various machine learning tasks in cosmology. The system effectively identifies and fixes errors in generated code, methodically improves performance through iterative refinement, and synthesizes new approaches by combining elements from successful implementations. A particularly notable aspect is the speed at which the system operates, completing entire research cycles in hours or days. For the Galaxy Zoo task, the system explored 50 implementation variations in approximately 72 hours, a breadth of experimentation that would require substantial human effort and time to match. Quantitative results show that models developed by the AI Cosmologist achieve performance comparable to baseline implementations created by human programmers. Moreover, the system's ability to explore diverse approaches occasionally leads to novel solutions that outperform conventional approaches, as evidenced by the Galaxy Zoo results exceeding the original Kaggle competition winner's performance.

The system's ability to learn from experimental results is particularly notable. Later iterations consistently show improvements over initial implementations, demonstrating effective transfer of knowledge across experimental runs. This progressive improvement suggests that the system is capable of accumulating insights and refining its approach based on empirical evidence, mirroring an important aspect of human scientific inquiry.

\subsection{Limitations}\label{subsec:limitations}

Despite its capabilities, the AI Cosmologist has several important limitations. While the system can effectively recombine existing approaches and implement known techniques, truly novel conceptual innovations remain challenging. The system operates primarily by adapting and refining established patterns rather than making fundamental breakthroughs in methodology. Additionally, the system requires well-specified problems and cannot yet formulate its own research questions, limiting its autonomy as a scientific agent.

The computational efficiency of generated code can vary and may not match the optimization level achieved by expert human programmers. This inefficiency can limit the scale of problems that can be practically addressed. Furthermore, the system lacks deep theoretical understanding that might guide more principled research approaches. It primarily learns from empirical results rather than from theoretical insights about the underlying physical processes.

An important limitation of the current work is that the examples demonstrated here involve datasets that are particularly well-suited for machine learning approaches. More exhaustive studies on more challenging and diverse datasets would be necessary to fully evaluate the system's capabilities and limitations across the spectrum of cosmological research problems.

\subsection{Future Directions}\label{subsec:future}

Future work could address these limitations through several promising avenues. Integration of theoretical knowledge bases could guide implementation choices with physical principles and established cosmological theory, potentially improving both the efficiency and scientific validity of generated solutions. Development of more sophisticated meta-learning capabilities could enhance the system's ability to transfer knowledge across different cosmological problems and datasets, accelerating learning in new domains.

Incorporation of human feedback and collaboration mechanisms would enable more effective human-AI teamwork, combining the complementary strengths of automated implementation with human scientific intuition. Extension to distributed systems would allow for larger-scale exploration of solution spaces, potentially enabling the discovery of more innovative approaches through broader experimentation. Additionally, expanding the system to handle multiple modalities of astronomical data simultaneously could increase its applicability to complex observational scenarios that combine imaging, spectroscopy, and time-series data.

Finally, developing frameworks to better evaluate and interpret the scientific significance of AI-generated results will be crucial for integrating systems like the AI Cosmologist into the broader scientific process. This includes tools for assessing the robustness, generalizability, and physical plausibility of automated findings, as well as methods for connecting machine-discovered patterns to theoretical understanding in cosmology.

\section{Conclusion}\label{sec:conclusion}
The AI Cosmologist represents a first step toward automating the cosmological data analysis and machine learning research process. By implementing a complete workflow from idea generation to experimental evaluation, the system demonstrates how AI can assist in or potentially automate significant portions of scientific discovery in machine learning.

While not yet capable of the creative leaps that characterize groundbreaking human research, the AI Cosmologist shows that methodical exploration, implementation, and iteration can be effectively performed by AI systems. This suggests a future where AI increasingly participates in its own development, with potentially profound implications for the pace and nature of progress in the field.

As capabilities improve, systems like AI Cosmologist may evolve from tools that automate routine aspects of research to collaborators that contribute novel insights and approaches. The  speed at which these systems operate—running dozens of experiments in parallel and completing in days what might take human researchers weeks or months—offers the potential to dramatically accelerate the pace of scientific discovery. This combination of speed and capability enables exploration of solution spaces at  scale, potentially uncovering valuable approaches that would remain undiscovered under traditional research timelines. This evolution promises to accelerate scientific progress while raising new questions about the changing role of human researchers in an increasingly automated scientific landscape.

\backmatter

\bmhead{Acknowledgments}
AM thanks colleagues for useful discussions involving the use of AI, including Steven Bamford, Ed Copeland, Simon Dye, Juan Garrahan, Anne Green, Maggie Lieu and Tony Padilla. 

\section*{Declarations}

\begin{itemize}
\item Funding: The work of A.M. was supported by an STFC Consolidated Grant [Grant No. ST/X000672/1]. For the purpose of open access, the author has applied a CC BY public copyright license to any Author Accepted Manuscript version arising.
\item Conflict of interest/Competing interests: The author declares no competing interests.
\item Ethics approval and consent to participate: Not applicable.
\item Consent for publication: Not applicable.
\item Data availability: The code and experimental data used in this paper are available on GitHub at \url{https://github.com/adammoss/aicosmologist}. The Galaxy Zoo 2 data is available via \url{https://www.kaggle.com/competitions/galaxy-zoo-the-galaxy-challenge/}. The Quijote simulation data is available via \url{https://quijote-simulations.readthedocs.io/en/latest/data_access.html}.
\item Materials availability: Not applicable.
\item Code availability: The code for the AI Cosmologist system and the experiments is available on GitHub at \url{https://github.com/adammoss/aicosmologist}.
\item Author contribution: A.M. conceived the study, developed the system, performed the experiments, analyzed the results, and wrote the manuscript.
\end{itemize}

\begin{appendices}

\section{Example Paper 1: Galaxy Zoo}\label{app:gz_paper}

\includepdf[pages=-]{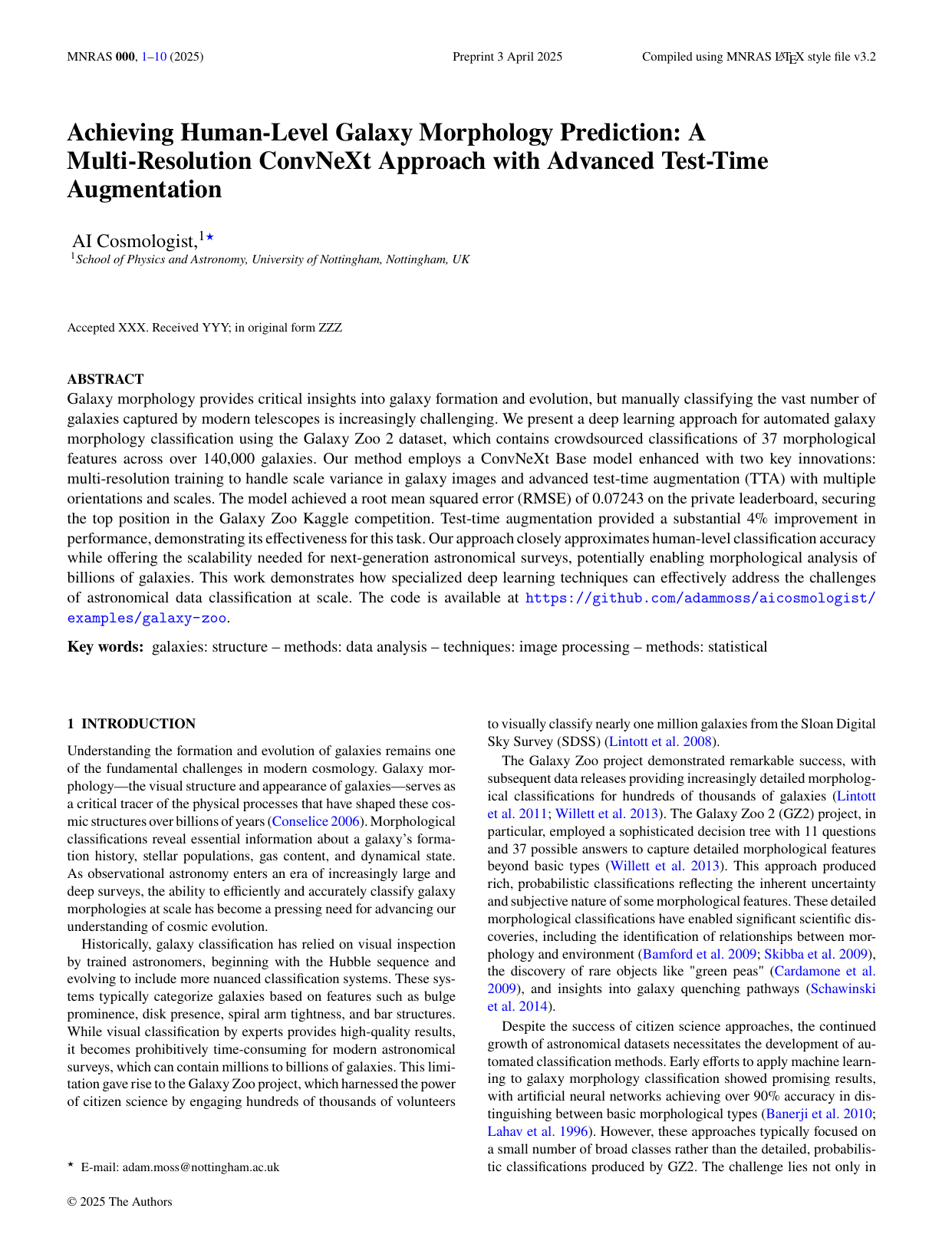}

\section{Example Paper 2: Quijote}\label{app:quijote_paper}

\includepdf[pages=-]{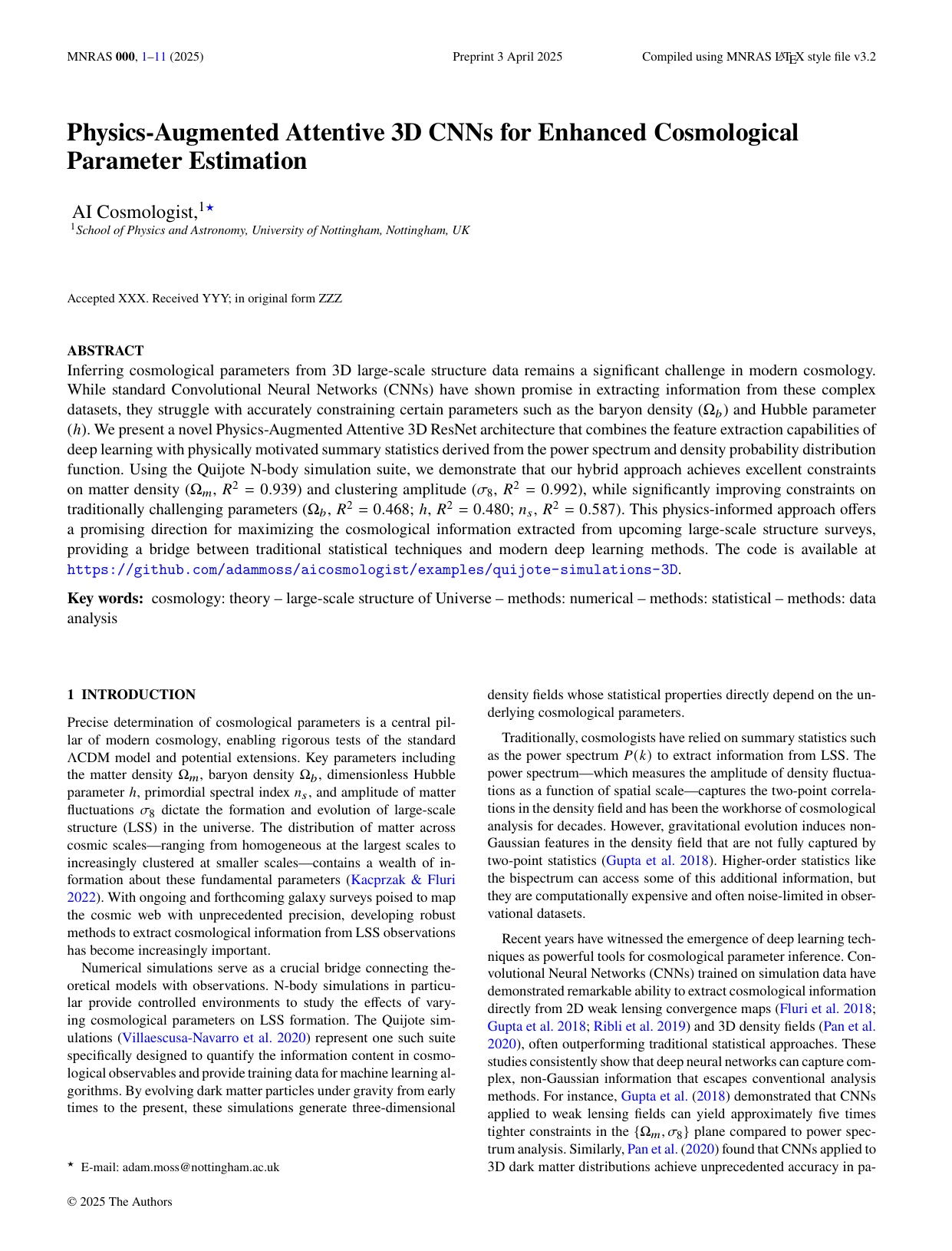}


\end{appendices}


\bibliography{references}


\begin{thebibliography}{47}
\ifx \bisbn   \undefined \def \bisbn  #1{ISBN #1}\fi
\ifx \binits  \undefined \def \binits#1{#1}\fi
\ifx \bauthor  \undefined \def \bauthor#1{#1}\fi
\ifx \batitle  \undefined \def \batitle#1{#1}\fi
\ifx \bjtitle  \undefined \def \bjtitle#1{#1}\fi
\ifx \bvolume  \undefined \def \bvolume#1{\textbf{#1}}\fi
\ifx \byear  \undefined \def \byear#1{#1}\fi
\ifx \bissue  \undefined \def \bissue#1{#1}\fi
\ifx \bfpage  \undefined \def \bfpage#1{#1}\fi
\ifx \blpage  \undefined \def \blpage #1{#1}\fi
\ifx \burl  \undefined \def \burl#1{\textsf{#1}}\fi
\ifx \doiurl  \undefined \def \doiurl#1{\url{https://doi.org/#1}}\fi
\ifx \betal  \undefined \def \betal{\textit{et al.}}\fi
\ifx \binstitute  \undefined \def \binstitute#1{#1}\fi
\ifx \binstitutionaled  \undefined \def \binstitutionaled#1{#1}\fi
\ifx \bctitle  \undefined \def \bctitle#1{#1}\fi
\ifx \beditor  \undefined \def \beditor#1{#1}\fi
\ifx \bpublisher  \undefined \def \bpublisher#1{#1}\fi
\ifx \bbtitle  \undefined \def \bbtitle#1{#1}\fi
\ifx \bedition  \undefined \def \bedition#1{#1}\fi
\ifx \bseriesno  \undefined \def \bseriesno#1{#1}\fi
\ifx \blocation  \undefined \def \blocation#1{#1}\fi
\ifx \bsertitle  \undefined \def \bsertitle#1{#1}\fi
\ifx \bsnm \undefined \def \bsnm#1{#1}\fi
\ifx \bsuffix \undefined \def \bsuffix#1{#1}\fi
\ifx \bparticle \undefined \def \bparticle#1{#1}\fi
\ifx \barticle \undefined \def \barticle#1{#1}\fi
\bibcommenthead
\ifx \bconfdate \undefined \def \bconfdate #1{#1}\fi
\ifx \botherref \undefined \def \botherref #1{#1}\fi
\ifx \url \undefined \def \url#1{\textsf{#1}}\fi
\ifx \bchapter \undefined \def \bchapter#1{#1}\fi
\ifx \bbook \undefined \def \bbook#1{#1}\fi
\ifx \bcomment \undefined \def \bcomment#1{#1}\fi
\ifx \oauthor \undefined \def \oauthor#1{#1}\fi
\ifx \citeauthoryear \undefined \def \citeauthoryear#1{#1}\fi
\ifx \endbibitem  \undefined \def \endbibitem {}\fi
\ifx \bconflocation  \undefined \def \bconflocation#1{#1}\fi
\ifx \arxivurl  \undefined \def \arxivurl#1{\textsf{#1}}\fi
\csname PreBibitemsHook\endcsname

\bibitem[\protect\citeauthoryear{Ivezi\'c et~al.}{2019}]{LSST:2008ijt}
\begin{barticle}
\bauthor{\bsnm{Ivezi\'c}, \binits{v.}}, \betal:
\batitle{{LSST: from Science Drivers to Reference Design and Anticipated Data Products}}.
\bjtitle{Astrophys. J.}
\bvolume{873}(\bissue{2}),
\bfpage{111}
(\byear{2019})
\doiurl{10.3847/1538-4357/ab042c}
{\href{https://arxiv.org/abs/0805.2366}{{arXiv:0805.2366}}}
{[astro-ph]}
\end{barticle}
\endbibitem

\bibitem[\protect\citeauthoryear{Laureijs et~al.}{2011}]{EUCLID:2011zbd}
\begin{botherref}
\oauthor{\bsnm{Laureijs}, \binits{R.}}, et al.:
{Euclid Definition Study Report}
(2011)
{\href{https://arxiv.org/abs/1110.3193}{{arXiv:1110.3193}}}
{[astro-ph.CO]}
\end{botherref}
\endbibitem

\bibitem[\protect\citeauthoryear{Bacon et~al.}{2020}]{SKA:2018ckk}
\begin{barticle}
\bauthor{\bsnm{Bacon}, \binits{D.J.}}, \betal:
\batitle{{Cosmology with Phase 1 of the Square Kilometre Array: Red Book 2018: Technical specifications and performance forecasts}}.
\bjtitle{Publ. Astron. Soc. Austral.}
\bvolume{37},
\bfpage{007}
(\byear{2020})
\doiurl{10.1017/pasa.2019.51}
{\href{https://arxiv.org/abs/1811.02743}{{arXiv:1811.02743}}}
{[astro-ph.CO]}
\end{barticle}
\endbibitem

\bibitem[\protect\citeauthoryear{Aghamousa et~al.}{2016}]{DESI:2016fyo}
\begin{botherref}
\oauthor{\bsnm{Aghamousa}, \binits{A.}}, et al.:
{The DESI Experiment Part I: Science,Targeting, and Survey Design}
(2016)
{\href{https://arxiv.org/abs/1611.00036}{{arXiv:1611.00036}}}
{[astro-ph.IM]}
\end{botherref}
\endbibitem

\bibitem[\protect\citeauthoryear{Villaescusa-Navarro et~al.}{2020}]{Villaescusa-Navarro:2019bje}
\begin{barticle}
\bauthor{\bsnm{Villaescusa-Navarro}, \binits{F.}}, \betal:
\batitle{{The Quijote simulations}}.
\bjtitle{Astrophys. J. Suppl.}
\bvolume{250}(\bissue{1}),
\bfpage{2}
(\byear{2020})
\doiurl{10.3847/1538-4365/ab9d82}
{\href{https://arxiv.org/abs/1909.05273}{{arXiv:1909.05273}}}
{[astro-ph.CO]}
\end{barticle}
\endbibitem

\bibitem[\protect\citeauthoryear{{Baron}}{2019}]{2019arXiv190407248B}
\begin{botherref}
\oauthor{\bsnm{{Baron}}, \binits{D.}}:
{Machine Learning in Astronomy: a practical overview}.
arXiv e-prints,
1904--07248
(2019)
\doiurl{10.48550/arXiv.1904.07248}
{\href{https://arxiv.org/abs/1904.07248}{{arXiv:1904.07248}}}
{[astro-ph.IM]}
\end{botherref}
\endbibitem

\bibitem[\protect\citeauthoryear{{Dieleman} et~al.}{2015}]{2015MNRAS.450.1441D}
\begin{barticle}
\bauthor{\bsnm{{Dieleman}}, \binits{S.}},
\bauthor{\bsnm{{Willett}}, \binits{K.W.}},
\bauthor{\bsnm{{Dambre}}, \binits{J.}}:
\batitle{{Rotation-invariant convolutional neural networks for galaxy morphology prediction}}.
\bjtitle{\mnras}
\bvolume{450}(\bissue{2}),
\bfpage{1441}--\blpage{1459}
(\byear{2015})
\doiurl{10.1093/mnras/stv632}
{\href{https://arxiv.org/abs/1503.07077}{{arXiv:1503.07077}}}
{[astro-ph.IM]}
\end{barticle}
\endbibitem

\bibitem[\protect\citeauthoryear{{Charnock} and {Moss}}{2017}]{2017ApJ...837L..28C}
\begin{barticle}
\bauthor{\bsnm{{Charnock}}, \binits{T.}},
\bauthor{\bsnm{{Moss}}, \binits{A.}}:
\batitle{{Deep Recurrent Neural Networks for Supernovae Classification}}.
\bjtitle{\apjl}
\bvolume{837}(\bissue{2}),
\bfpage{28}
(\byear{2017})
\doiurl{10.3847/2041-8213/aa603d}
{\href{https://arxiv.org/abs/1606.07442}{{arXiv:1606.07442}}}
{[astro-ph.IM]}
\end{barticle}
\endbibitem

\bibitem[\protect\citeauthoryear{{Bloom} et~al.}{2012}]{2012PASP..124.1175B}
\begin{barticle}
\bauthor{\bsnm{{Bloom}}, \binits{J.S.}},
\bauthor{\bsnm{{Richards}}, \binits{J.W.}},
\bauthor{\bsnm{{Nugent}}, \binits{P.E.}},
\bauthor{\bsnm{{Quimby}}, \binits{R.M.}},
\bauthor{\bsnm{{Kasliwal}}, \binits{M.M.}},
\bauthor{\bsnm{{Starr}}, \binits{D.L.}},
\bauthor{\bsnm{{Poznanski}}, \binits{D.}},
\bauthor{\bsnm{{Ofek}}, \binits{E.O.}},
\bauthor{\bsnm{{Cenko}}, \binits{S.B.}},
\bauthor{\bsnm{{Butler}}, \binits{N.R.}},
\bauthor{\bsnm{{Kulkarni}}, \binits{S.R.}},
\bauthor{\bsnm{{Gal-Yam}}, \binits{A.}},
\bauthor{\bsnm{{Law}}, \binits{N.}}:
\batitle{{Automating Discovery and Classification of Transients and Variable Stars in the Synoptic Survey Era}}.
\bjtitle{\pasp}
\bvolume{124}(\bissue{921}),
\bfpage{1175}
(\byear{2012})
\doiurl{10.1086/668468}
{\href{https://arxiv.org/abs/1106.5491}{{arXiv:1106.5491}}}
{[astro-ph.IM]}
\end{barticle}
\endbibitem

\bibitem[\protect\citeauthoryear{{Petrillo} et~al.}{2017}]{2017MNRAS.472.1129P}
\begin{barticle}
\bauthor{\bsnm{{Petrillo}}, \binits{C.E.}},
\bauthor{\bsnm{{Tortora}}, \binits{C.}},
\bauthor{\bsnm{{Chatterjee}}, \binits{S.}},
\bauthor{\bsnm{{Vernardos}}, \binits{G.}},
\bauthor{\bsnm{{Koopmans}}, \binits{L.V.E.}},
\bauthor{\bsnm{{Verdoes Kleijn}}, \binits{G.}},
\bauthor{\bsnm{{Napolitano}}, \binits{N.R.}},
\bauthor{\bsnm{{Covone}}, \binits{G.}},
\bauthor{\bsnm{{Schneider}}, \binits{P.}},
\bauthor{\bsnm{{Grado}}, \binits{A.}},
\bauthor{\bsnm{{McFarland}}, \binits{J.}}:
\batitle{{Finding strong gravitational lenses in the Kilo Degree Survey with Convolutional Neural Networks}}.
\bjtitle{\mnras}
\bvolume{472}(\bissue{1}),
\bfpage{1129}--\blpage{1150}
(\byear{2017})
\doiurl{10.1093/mnras/stx2052}
{\href{https://arxiv.org/abs/1702.07675}{{arXiv:1702.07675}}}
{[astro-ph.GA]}
\end{barticle}
\endbibitem

\bibitem[\protect\citeauthoryear{{Collister} and {Lahav}}{2004}]{2004PASP..116..345C}
\begin{barticle}
\bauthor{\bsnm{{Collister}}, \binits{A.A.}},
\bauthor{\bsnm{{Lahav}}, \binits{O.}}:
\batitle{{ANNz: Estimating Photometric Redshifts Using Artificial Neural Networks}}.
\bjtitle{\pasp}
\bvolume{116}(\bissue{818}),
\bfpage{345}--\blpage{351}
(\byear{2004})
\doiurl{10.1086/383254}
{\href{https://arxiv.org/abs/astro-ph/0311058}{{arXiv:astro-ph/0311058}}}
{[astro-ph]}
\end{barticle}
\endbibitem

\bibitem[\protect\citeauthoryear{{George} and {Huerta}}{2018}]{2018PhRvD..97d4039G}
\begin{barticle}
\bauthor{\bsnm{{George}}, \binits{D.}},
\bauthor{\bsnm{{Huerta}}, \binits{E.A.}}:
\batitle{{Deep neural networks to enable real-time multimessenger astrophysics}}.
\bjtitle{\prd}
\bvolume{97}(\bissue{4}),
\bfpage{044039}
(\byear{2018})
\doiurl{10.1103/PhysRevD.97.044039}
{\href{https://arxiv.org/abs/1701.00008}{{arXiv:1701.00008}}}
{[astro-ph.IM]}
\end{barticle}
\endbibitem

\bibitem[\protect\citeauthoryear{Butler et~al.}{2018}]{Butler2018MLMaterials}
\begin{barticle}
\bauthor{\bsnm{Butler}, \binits{K.T.}},
\bauthor{\bsnm{Davies}, \binits{D.W.}},
\bauthor{\bsnm{Cartwright}, \binits{H.}},
\bauthor{\bsnm{Isayev}, \binits{O.}},
\bauthor{\bsnm{Walsh}, \binits{A.}}:
\batitle{{Machine learning for molecular and materials science}}.
\bjtitle{Nature}
\bvolume{559}(\bissue{7715}),
\bfpage{547}--\blpage{555}
(\byear{2018})
\doiurl{10.1038/s41586-018-0337-2}
\end{barticle}
\endbibitem

\bibitem[\protect\citeauthoryear{Jumper et~al.}{2021}]{Jumper2021AlphaFold}
\begin{barticle}
\bauthor{\bsnm{Jumper}, \binits{J.}},
\bauthor{\bsnm{Evans}, \binits{R.}},
\bauthor{\bsnm{Pritzel}, \binits{A.}},
\bauthor{\bsnm{Green}, \binits{T.}},
\bauthor{\bsnm{Figurnov}, \binits{S.}},
\bauthor{\bsnm{Ronneberger}, \binits{O.}},
\bauthor{\bsnm{Tunyasuvunakool}, \binits{K.}},
\bauthor{\bsnm{Bates}, \binits{R.}},
\bauthor{\bsnm{{\ldots}}},
\bauthor{\bsnm{Hassabis}, \binits{D.}}:
\batitle{{Highly accurate protein structure prediction with AlphaFold}}.
\bjtitle{Nature}
\bvolume{596}(\bissue{7873}),
\bfpage{583}--\blpage{589}
(\byear{2021})
\doiurl{10.1038/s41586-021-03819-2}
\end{barticle}
\endbibitem

\bibitem[\protect\citeauthoryear{Reichstein et~al.}{2019}]{Reichstein2019ESDLearning}
\begin{barticle}
\bauthor{\bsnm{Reichstein}, \binits{M.}},
\bauthor{\bsnm{Camps-Valls}, \binits{G.}},
\bauthor{\bsnm{Stevens}, \binits{B.}},
\bauthor{\bsnm{Jung}, \binits{M.}},
\bauthor{\bsnm{Denzler}, \binits{J.}},
\bauthor{\bsnm{Carvalhais}, \binits{N.}},
\bauthor{\bsnm{Prabhat}}:
\batitle{{Deep learning and process understanding for data-driven Earth system science}}.
\bjtitle{Nature}
\bvolume{566}(\bissue{7743}),
\bfpage{195}--\blpage{204}
(\byear{2019})
\doiurl{10.1038/s41586-019-0912-1}
\end{barticle}
\endbibitem

\bibitem[\protect\citeauthoryear{Z{\"o}ller and Huber}{2021}]{zoller2021benchmark}
\begin{barticle}
\bauthor{\bsnm{Z{\"o}ller}, \binits{M.-A.}},
\bauthor{\bsnm{Huber}, \binits{M.F.}}:
\batitle{Benchmark and survey of automated machine learning frameworks}.
\bjtitle{Journal of artificial intelligence research}
\bvolume{70},
\bfpage{409}--\blpage{472}
(\byear{2021})
\end{barticle}
\endbibitem

\bibitem[\protect\citeauthoryear{Chen and et~al.}{2021}]{Chen2021Codex}
\begin{botherref}
\oauthor{\bsnm{Chen}, \binits{M.}},
\oauthor{\bsnm{al.}}:
Evaluating large language models trained on code.
arXiv preprint arXiv:2107.03374
(2021)
\end{botherref}
\endbibitem

\bibitem[\protect\citeauthoryear{Li and et~al.}{2022}]{Li2022AlphaCode}
\begin{barticle}
\bauthor{\bsnm{Li}, \binits{Y.}},
\bauthor{\bsnm{al.}}:
\batitle{Competition-level code generation with alphacode}.
\bjtitle{Science}
\bvolume{378}(\bissue{6624}),
\bfpage{1092}--\blpage{1100}
(\byear{2022})
\end{barticle}
\endbibitem

\bibitem[\protect\citeauthoryear{Yao et~al.}{2023}]{Yao2023ReAct}
\begin{bchapter}
\bauthor{\bsnm{Yao}, \binits{S.}},
\bauthor{\bsnm{Zhao}, \binits{J.}},
\bauthor{\bsnm{Yu}, \binits{D.}},
\bauthor{\bsnm{Du}, \binits{N.}},
\bauthor{\bsnm{Shafran}, \binits{I.}},
\bauthor{\bsnm{Narasimhan}, \binits{K.}},
\bauthor{\bsnm{Cao}, \binits{Y.}}:
\bctitle{{ReAct: Synergizing Reasoning and Acting in Language Models}}.
In: \bbtitle{Proc. International Conference on Learning Representations (ICLR)}
(\byear{2023})
\end{bchapter}
\endbibitem

\bibitem[\protect\citeauthoryear{Shinn and et~al.}{2023}]{Shinn2023Reflexion}
\begin{botherref}
\oauthor{\bsnm{Shinn}, \binits{N.}},
\oauthor{\bsnm{al.}}:
Reflexion: Language agents with verbal reinforcement learning.
arXiv preprint arXiv:2303.11366
(2023)
\end{botherref}
\endbibitem

\bibitem[\protect\citeauthoryear{Yang et~al.}{2024}]{yang2024swe}
\begin{barticle}
\bauthor{\bsnm{Yang}, \binits{J.}},
\bauthor{\bsnm{Jimenez}, \binits{C.}},
\bauthor{\bsnm{Wettig}, \binits{A.}},
\bauthor{\bsnm{Lieret}, \binits{K.}},
\bauthor{\bsnm{Yao}, \binits{S.}},
\bauthor{\bsnm{Narasimhan}, \binits{K.}},
\bauthor{\bsnm{Press}, \binits{O.}}:
\batitle{Swe-agent: Agent-computer interfaces enable automated software engineering}.
\bjtitle{Advances in Neural Information Processing Systems}
\bvolume{37},
\bfpage{50528}--\blpage{50652}
(\byear{2024})
\end{barticle}
\endbibitem

\bibitem[\protect\citeauthoryear{M.~Bran et~al.}{2024}]{m2024augmenting}
\begin{barticle}
\bauthor{\bsnm{M.~Bran}, \binits{A.}},
\bauthor{\bsnm{Cox}, \binits{S.}},
\bauthor{\bsnm{Schilter}, \binits{O.}},
\bauthor{\bsnm{Baldassari}, \binits{C.}},
\bauthor{\bsnm{White}, \binits{A.D.}},
\bauthor{\bsnm{Schwaller}, \binits{P.}}:
\batitle{Augmenting large language models with chemistry tools}.
\bjtitle{Nature Machine Intelligence}
\bvolume{6}(\bissue{5}),
\bfpage{525}--\blpage{535}
(\byear{2024})
\end{barticle}
\endbibitem

\bibitem[\protect\citeauthoryear{Hutter et~al.}{2019}]{Hutter2019}
\begin{botherref}
\oauthor{\bsnm{Hutter}, \binits{F.}},
\oauthor{\bsnm{Kotthoff}, \binits{L.}},
\oauthor{\bsnm{Vanschoren}, \binits{J.}}:
Automated machine learning: Methods, systems, challenges.
Automated Machine Learning
(2019)
\end{botherref}
\endbibitem

\bibitem[\protect\citeauthoryear{Elsken et~al.}{2019}]{Elsken2019}
\begin{barticle}
\bauthor{\bsnm{Elsken}, \binits{T.}},
\bauthor{\bsnm{Metzen}, \binits{J.H.}},
\bauthor{\bsnm{Hutter}, \binits{F.}}:
\batitle{Neural architecture search: A survey}.
\bjtitle{Journal of Machine Learning Research}
\bvolume{20}(\bissue{55}),
\bfpage{1}--\blpage{21}
(\byear{2019})
\end{barticle}
\endbibitem

\bibitem[\protect\citeauthoryear{Thornton et~al.}{2013}]{Thornton2013}
\begin{bchapter}
\bauthor{\bsnm{Thornton}, \binits{C.}},
\bauthor{\bsnm{Hutter}, \binits{F.}},
\bauthor{\bsnm{Hoos}, \binits{H.H.}},
\bauthor{\bsnm{Leyton-Brown}, \binits{K.}}:
\bctitle{Auto-weka: Combined selection and hyperparameter optimization of classification algorithms}.
In: \bbtitle{KDD},
pp. \bfpage{847}--\blpage{855}
(\byear{2013})
\end{bchapter}
\endbibitem

\bibitem[\protect\citeauthoryear{Feurer and et~al.}{2015}]{Feurer2015}
\begin{bchapter}
\bauthor{\bsnm{Feurer}, \binits{M.}},
\bauthor{\bsnm{al.}}:
\bctitle{Efficient and robust automated machine learning}.
In: \bbtitle{NIPS},
pp. \bfpage{2962}--\blpage{2970}
(\byear{2015})
\end{bchapter}
\endbibitem

\bibitem[\protect\citeauthoryear{Vanschoren}{2019}]{Vanschoren2019}
\begin{botherref}
\oauthor{\bsnm{Vanschoren}, \binits{J.}}:
Meta-learning: A survey.
arXiv preprint arXiv:1810.03548
(2019)
\end{botherref}
\endbibitem

\bibitem[\protect\citeauthoryear{Olson et~al.}{2016}]{Olson2016}
\begin{bchapter}
\bauthor{\bsnm{Olson}, \binits{R.S.}},
\bauthor{\bsnm{Bartley}, \binits{N.}},
\bauthor{\bsnm{Urbanowicz}, \binits{R.J.}},
\bauthor{\bsnm{Moore}, \binits{J.H.}}:
\bctitle{Evaluation of a tree-based pipeline optimization tool for automating data science}.
In: \bbtitle{Genetic and Evolutionary Computation Conference (GECCO) Companion},
pp. \bfpage{485}--\blpage{492}
(\byear{2016})
\end{bchapter}
\endbibitem

\bibitem[\protect\citeauthoryear{Zoph and Le}{2017}]{Zoph2017}
\begin{bchapter}
\bauthor{\bsnm{Zoph}, \binits{B.}},
\bauthor{\bsnm{Le}, \binits{Q.V.}}:
\bctitle{Neural architecture search with reinforcement learning}.
In: \bbtitle{ICLR}
(\byear{2017})
\end{bchapter}
\endbibitem

\bibitem[\protect\citeauthoryear{Real et~al.}{2019}]{Real2019}
\begin{bchapter}
\bauthor{\bsnm{Real}, \binits{E.}},
\bauthor{\bsnm{Aggarwal}, \binits{A.}},
\bauthor{\bsnm{Huang}, \binits{Y.}},
\bauthor{\bsnm{Le}, \binits{Q.V.}}:
\bctitle{Regularized evolution for image classifier architecture search}.
In: \bbtitle{Proceedings of the AAAI Conference on Artificial Intelligence},
vol. \bseriesno{33},
pp. \bfpage{4780}--\blpage{4789}
(\byear{2019})
\end{bchapter}
\endbibitem

\bibitem[\protect\citeauthoryear{Liu et~al.}{2019}]{Liu2019}
\begin{bchapter}
\bauthor{\bsnm{Liu}, \binits{H.}},
\bauthor{\bsnm{Simonyan}, \binits{K.}},
\bauthor{\bsnm{Yang}, \binits{Y.}}:
\bctitle{{DARTS}: Differentiable architecture search}.
In: \bbtitle{International Conference on Learning Representations (ICLR)}
(\byear{2019})
\end{bchapter}
\endbibitem

\bibitem[\protect\citeauthoryear{Finn et~al.}{2017}]{Finn2017}
\begin{bchapter}
\bauthor{\bsnm{Finn}, \binits{C.}},
\bauthor{\bsnm{Abbeel}, \binits{P.}},
\bauthor{\bsnm{Levine}, \binits{S.}}:
\bctitle{Model-agnostic meta-learning for fast adaptation of deep networks}.
In: \bbtitle{Proceedings of the 34th International Conference on Machine Learning (ICML)},
pp. \bfpage{1126}--\blpage{1135}
(\byear{2017})
\end{bchapter}
\endbibitem

\bibitem[\protect\citeauthoryear{Kruk et~al.}{2022}]{Kruk2022}
\begin{barticle}
\bauthor{\bsnm{Kruk}, \binits{S.J.}},
\bauthor{\bsnm{Garc{\'i}a~Mart{\'i}n}, \binits{P.}},
\bauthor{\bsnm{Popescu}, \binits{M.}},
\bauthor{\bsnm{Mer{\'i}n}, \binits{B.}},
\bauthor{\bsnm{Mahlke}, \binits{M.}},
\bauthor{\bsnm{Carry}, \binits{B.}},
\bauthor{\bsnm{Thomson}, \binits{R.}},
\bauthor{\bsnm{Karada{\u g}}, \binits{S.}},
\bauthor{\bsnm{Dur{\'a}n}, \binits{J.}},
\bauthor{\bsnm{al.}}:
\batitle{{Hubble Asteroid Hunter. I. Identifying asteroid trails in HST images}}.
\bjtitle{Astronomy \& Astrophysics}
\bvolume{661},
\bfpage{85}
(\byear{2022})
\end{barticle}
\endbibitem

\bibitem[\protect\citeauthoryear{Tarsitano et~al.}{2022}]{Tarsitano2022}
\begin{barticle}
\bauthor{\bsnm{Tarsitano}, \binits{C.}},
\bauthor{\bsnm{La~Barbera}, \binits{F.}},
\bauthor{\bsnm{Vavilova}, \binits{I.}},
\bauthor{\bsnm{al.}}:
\batitle{Image feature extraction and galaxy classification: a novel approach using automated machine learning}.
\bjtitle{Monthly Notices of the Royal Astronomical Society}
\bvolume{511},
\bfpage{3330}--\blpage{3348}
(\byear{2022})
\end{barticle}
\endbibitem

\bibitem[\protect\citeauthoryear{Chen et~al.}{2021}]{Chen2021}
\begin{botherref}
\oauthor{\bsnm{Chen}, \binits{M.}},
\oauthor{\bsnm{Tworek}, \binits{J.}},
\oauthor{\bsnm{Jun}, \binits{H.}},
\oauthor{\bsnm{Yuan}, \binits{Q.}},
\oauthor{\bsnm{Pinto}, \binits{H.P.d.O.}},
\oauthor{\bsnm{Kaplan}, \binits{J.}},
\oauthor{\bsnm{Edwards}, \binits{H.}},
\oauthor{\bsnm{Burda}, \binits{Y.}},
\oauthor{\bsnm{Joseph}, \binits{N.}},
\oauthor{\bsnm{Brockman}, \binits{G.}},
\oauthor{\bsnm{Ray}, \binits{A.}},
\oauthor{\bsnm{Puri}, \binits{R.}},
\oauthor{\bsnm{Krueger}, \binits{G.}},
\oauthor{\bsnm{Petrov}, \binits{M.}},
\oauthor{\bsnm{Khlaaf}, \binits{H.}},
\oauthor{\bsnm{Sastry}, \binits{G.}},
\oauthor{\bsnm{Mishkin}, \binits{P.}},
\oauthor{\bsnm{Chan}, \binits{B.}},
\oauthor{\bsnm{Gray}, \binits{S.}},
\oauthor{\bsnm{Ryder}, \binits{N.}},
\oauthor{\bsnm{Pavlov}, \binits{M.}},
\oauthor{\bsnm{Power}, \binits{A.}},
\oauthor{\bsnm{Kaiser}, \binits{L.}},
\oauthor{\bsnm{Bavarian}, \binits{M.}},
\oauthor{\bsnm{Winter}, \binits{C.}},
\oauthor{\bsnm{Tillet}, \binits{P.}},
\oauthor{\bsnm{Such}, \binits{F.P.}},
\oauthor{\bsnm{Cummings}, \binits{D.}},
\oauthor{\bsnm{Plappert}, \binits{M.}},
\oauthor{\bsnm{Chantzis}, \binits{F.}},
\oauthor{\bsnm{Barnes}, \binits{E.}},
\oauthor{\bsnm{Herbert-Voss}, \binits{A.}},
\oauthor{\bsnm{Guss}, \binits{W.H.}},
\oauthor{\bsnm{Nichol}, \binits{A.}},
\oauthor{\bsnm{Paino}, \binits{A.}},
\oauthor{\bsnm{Tezak}, \binits{N.}},
\oauthor{\bsnm{Tang}, \binits{J.}},
\oauthor{\bsnm{Babuschkin}, \binits{I.}},
\oauthor{\bsnm{Balaji}, \binits{S.}},
\oauthor{\bsnm{Jain}, \binits{S.}},
\oauthor{\bsnm{Saunders}, \binits{W.}},
\oauthor{\bsnm{Hesse}, \binits{C.}},
\oauthor{\bsnm{Carr}, \binits{A.N.}},
\oauthor{\bsnm{Leike}, \binits{J.}},
\oauthor{\bsnm{Achiam}, \binits{J.}},
\oauthor{\bsnm{Misra}, \binits{V.}},
\oauthor{\bsnm{Morikawa}, \binits{E.}},
\oauthor{\bsnm{Radford}, \binits{A.}},
\oauthor{\bsnm{Knight}, \binits{M.}},
\oauthor{\bsnm{Brundage}, \binits{M.}},
\oauthor{\bsnm{Murati}, \binits{M.}},
\oauthor{\bsnm{Mayer}, \binits{K.}},
\oauthor{\bsnm{Welinder}, \binits{P.}},
\oauthor{\bsnm{McGrew}, \binits{B.}},
\oauthor{\bsnm{Amodei}, \binits{D.}},
\oauthor{\bsnm{McCandlish}, \binits{S.}},
\oauthor{\bsnm{Sutskever}, \binits{I.}},
\oauthor{\bsnm{Zaremba}, \binits{W.}}:
Evaluating large language models trained on code.
arXiv preprint arXiv:2107.03374
(2021)
\end{botherref}
\endbibitem

\bibitem[\protect\citeauthoryear{Nijkamp et~al.}{2022}]{Nijkamp2022}
\begin{botherref}
\oauthor{\bsnm{Nijkamp}, \binits{E.}},
\oauthor{\bsnm{Pang}, \binits{B.}},
\oauthor{\bsnm{Hayashi}, \binits{H.}},
\oauthor{\bsnm{Tu}, \binits{L.}},
\oauthor{\bsnm{Wang}, \binits{H.}},
\oauthor{\bsnm{Zhou}, \binits{Y.}},
\oauthor{\bsnm{Savarese}, \binits{S.}},
\oauthor{\bsnm{Xiong}, \binits{C.}}:
{CodeGen}: An Open Large Language Model for Code with Multi-Turn Program Synthesis.
arXiv preprint arXiv:2203.13474
(2022)
\end{botherref}
\endbibitem

\bibitem[\protect\citeauthoryear{Fried et~al.}{2023}]{Fried2023}
\begin{bchapter}
\bauthor{\bsnm{Fried}, \binits{D.}},
\bauthor{\bsnm{Aghajanyan}, \binits{A.}},
\bauthor{\bsnm{Lin}, \binits{J.}},
\bauthor{\bsnm{Wang}, \binits{S.I.}},
\bauthor{\bsnm{Wallace}, \binits{E.}},
\bauthor{\bsnm{Shi}, \binits{F.}},
\bauthor{\bsnm{Zhong}, \binits{R.}},
\bauthor{\bsnm{Yih}, \binits{W.-t.}},
\bauthor{\bsnm{Zettlemoyer}, \binits{L.}},
\bauthor{\bsnm{Lewis}, \binits{M.}}:
\bctitle{{InCoder}: A generative model for code infilling and synthesis}.
In: \bbtitle{International Conference on Learning Representations (ICLR)}
(\byear{2023})
\end{bchapter}
\endbibitem

\bibitem[\protect\citeauthoryear{Li et~al.}{2023}]{Li2023StarCoder}
\begin{botherref}
\oauthor{\bsnm{Li}, \binits{R.}},
\oauthor{\bsnm{Allal}, \binits{L.B.}},
\oauthor{\bsnm{Zi}, \binits{Y.}},
\oauthor{\bsnm{Muennighoff}, \binits{N.}},
\oauthor{\bsnm{Kocetkov}, \binits{D.}},
\oauthor{\bsnm{Mou}, \binits{C.}},
\oauthor{\bsnm{Marone}, \binits{M.}},
\oauthor{\bsnm{Akiki}, \binits{C.}}, et al.:
{StarCoder}: May the source be with you!
arXiv preprint arXiv:2305.06161
(2023)
\end{botherref}
\endbibitem

\bibitem[\protect\citeauthoryear{Yao et~al.}{2023}]{Yao2023}
\begin{botherref}
\oauthor{\bsnm{Yao}, \binits{S.}},
\oauthor{\bsnm{Zhao}, \binits{J.}},
\oauthor{\bsnm{Yu}, \binits{D.}},
\oauthor{\bsnm{Du}, \binits{N.}},
\oauthor{\bsnm{Shafran}, \binits{I.}},
\oauthor{\bsnm{Narasimhan}, \binits{K.}},
\oauthor{\bsnm{Cao}, \binits{Y.}}:
{ReAct}: Synergizing reasoning and acting in language models.
Proceedings of the 40th International Conference on Machine Learning (ICML),
11837--11852
(2023)
\end{botherref}
\endbibitem

\bibitem[\protect\citeauthoryear{Shinn et~al.}{2023}]{Shinn2023}
\begin{bchapter}
\bauthor{\bsnm{Shinn}, \binits{N.}},
\bauthor{\bsnm{Cassano}, \binits{F.}},
\bauthor{\bsnm{Berman}, \binits{E.}},
\bauthor{\bsnm{Gopinath}, \binits{A.}},
\bauthor{\bsnm{Narasimhan}, \binits{K.}},
\bauthor{\bsnm{Yao}, \binits{S.}}:
\bctitle{Reflexion: Language agents with verbal reinforcement learning}.
In: \bbtitle{Advances in Neural Information Processing Systems 36 (NeurIPS 2023)},
pp. \bfpage{41895}--\blpage{41908}
(\byear{2023})
\end{bchapter}
\endbibitem

\bibitem[\protect\citeauthoryear{Shen et~al.}{2023}]{Shen2023}
\begin{botherref}
\oauthor{\bsnm{Shen}, \binits{Y.}},
\oauthor{\bsnm{Song}, \binits{K.}},
\oauthor{\bsnm{Tan}, \binits{X.}},
\oauthor{\bsnm{Li}, \binits{D.}},
\oauthor{\bsnm{Lu}, \binits{W.}},
\oauthor{\bsnm{Zhuang}, \binits{Y.}}:
{HuggingGPT}: Solving AI Tasks with ChatGPT and its Friends in Hugging Face.
arXiv preprint arXiv:2303.17580
(2023)
\end{botherref}
\endbibitem

\bibitem[\protect\citeauthoryear{Boiko et~al.}{2023}]{Boiko2023}
\begin{barticle}
\bauthor{\bsnm{Boiko}, \binits{D.A.}},
\bauthor{\bsnm{MacKnight}, \binits{R.}},
\bauthor{\bsnm{Kline}, \binits{B.}},
\bauthor{\bsnm{Gomes}, \binits{G.}}:
\batitle{Autonomous chemical research with large language models}.
\bjtitle{Nature}
\bvolume{624},
\bfpage{570}--\blpage{578}
(\byear{2023})
\end{barticle}
\endbibitem

\bibitem[\protect\citeauthoryear{Cranmer et~al.}{2020}]{Cranmer2020}
\begin{barticle}
\bauthor{\bsnm{Cranmer}, \binits{K.}},
\bauthor{\bsnm{Brehmer}, \binits{J.}},
\bauthor{\bsnm{Louppe}, \binits{G.}}:
\batitle{The frontier of simulation-based inference}.
\bjtitle{PNAS}
\bvolume{117}(\bissue{48}),
\bfpage{30055}--\blpage{30062}
(\byear{2020})
\end{barticle}
\endbibitem

\bibitem[\protect\citeauthoryear{Brehmer and et~al.}{2020}]{Brehmer2019}
\begin{barticle}
\bauthor{\bsnm{Brehmer}, \binits{J.}},
\bauthor{\bsnm{al.}}:
\batitle{Mining gold from implicit models to improve likelihood-free inference}.
\bjtitle{PNAS}
\bvolume{117}(\bissue{10}),
\bfpage{5242}--\blpage{5249}
(\byear{2020})
\end{barticle}
\endbibitem

\bibitem[\protect\citeauthoryear{Xu and et~al.}{2023}]{Xu2023ExpertIter}
\begin{botherref}
\oauthor{\bsnm{Xu}, \binits{M.}},
\oauthor{\bsnm{al.}}:
Expert iteration with language models for scientific discovery.
arXiv preprint arXiv:2303.03494
(2023)
\end{botherref}
\endbibitem

\bibitem[\protect\citeauthoryear{Laverick et~al.}{2024}]{Laverick:2024fyh}
\begin{botherref}
\oauthor{\bsnm{Laverick}, \binits{A.}},
\oauthor{\bsnm{Surrao}, \binits{K.}},
\oauthor{\bsnm{Zubeldia}, \binits{I.}},
\oauthor{\bsnm{Bolliet}, \binits{B.}},
\oauthor{\bsnm{Cranmer}, \binits{M.}},
\oauthor{\bsnm{Lewis}, \binits{A.}},
\oauthor{\bsnm{Sherwin}, \binits{B.}},
\oauthor{\bsnm{Lesgourgues}, \binits{J.}}:
{Multi-Agent System for Cosmological Parameter Analysis}
(2024)
\end{botherref}
\endbibitem

\bibitem[\protect\citeauthoryear{Willett et~al.}{2013}]{Willett:2013wda}
\begin{barticle}
\bauthor{\bsnm{Willett}, \binits{K.W.}}, \betal:
\batitle{{Galaxy Zoo 2: detailed morphological classifications for 304,122 galaxies from the Sloan Digital Sky Survey}}.
\bjtitle{Mon. Not. Roy. Astron. Soc.}
\bvolume{435},
\bfpage{2835}
(\byear{2013})
\doiurl{10.1093/mnras/stt1458}
{\href{https://arxiv.org/abs/1308.3496}{{arXiv:1308.3496}}}
{[astro-ph.CO]}
\end{barticle}
\endbibitem

\end{thebibliography}

\end{document}